\documentclass[12pt,a4paper]{article}

\usepackage{graphicx}
\usepackage{amsmath,amssymb,amsfonts,amsthm}
\usepackage{braket}
\usepackage{array}
\usepackage{comment}
\usepackage[dvipdfmx]{color}
\usepackage{algorithmic,algorithm}
\usepackage{bm}
\usepackage{multirow,array}
\usepackage{url}

\newtheorem{Theorem}{Theorem}
\newtheorem{Definition}[Theorem]{Definition}

\newtheorem{Lemma}[Theorem]{Lemma}
\newtheorem{Corollary}[Theorem]{Corollary}
\newtheorem{Remark}[Theorem]{Remark}
\newtheorem{Fact}[Theorem]{Fact}
\newtheorem{Example}[Theorem]{Example}

\newcommand{\cratio}{$\text{(C1)}^{-}$}
\newcommand{\coutd}{$\text{(C2)}^{-}$}
\newcommand{\cind}{$\text{(C3)}^{-}$}

\begin{document}
\title{Quantum Deletion Codes derived from Classical Deletion Codes (Extended Abstract)}

\author{
Manabu HAGIWARA
\thanks{
Department of Mathematics and Informatics,
Graduate School of Science,
Chiba University
1-33 Yayoi-cho, Inage-ku, Chiba City,
Chiba Pref., JAPAN, 263-0022.
E-mail: hagiwara@math.s.chiba-u.ac.jp
}
}

\date{2022/08/11}

\maketitle

\begin{abstract}
This manuscript is an extended abstract version of the paper 
entitled ``Quantum Deletion Codes derived from Classical Deletion Codes.''
The paper contributes to the fundamental theory for quantum deletion error-correcting codes.
The paper proposes a code construction condition
for a partition of classical deletion error-correcting codes to derive quantum deletion error-correcting codes.
The construction methods in this paper
give examples of quantum codes that can correct single-quantum deletion errors and 
have a code rate arbitrarily close to 1, while the previously known quantum deletion code rates are close to 0 for long length.
This manuscript omits the proofs of the statements in the paper.
\end{abstract}

\section{Introduction}
This manuscript is an extended abstract version of the paper 
entitled ``Quantum Deletion Codes derived from Classical Deletion Codes.''
The paper aims to construct error-correcting codes for quantum deletion errors and to develop a part of the fundamental theory for this purpose.
In particular we deal with single deletion errors.
Multiple deletion error-correction seems too challenging to deal with it at this stage, and we leave it to future research.

In classical coding theory, deletion error-correcting codes have been studied since the 1960s \cite{levenshtein1966binary}.
An error model called deletion error was proposed as one of the errors that occur in communication synchronization.
Levenshtein is the proposer and a pioneer in classical deletion error-correction coding theory.
He proved that the codes called VT codes are classical single-deletion error-correcting codes.
He also showed that deletion error-correcting codes could also correct other errors, e.g., insertion errors and substitution errors.
With the recent development of information technology, deletion errors have begun to be pointed out as errors on Racetrack Memory/Domain Wall Motion Memory \cite{chee2018coding,mappouras2019greenflag,ollivier2019leveraging,vahid2017correcting}, and DNA Storage \cite{antkowiak2020low,buschmann2013levenshtein,xue2020construction}.
Accordingly, results on deletion codes have been published from various researchers.
In other words, the diversification of classical information technology has increased the value of deletion error-correcting codes.
The author expects a similar phenomenon to happen in quantum information theory, in particular, quantum deletion error-correction.

Although classical single-deletion error-correcting codes were discovered in the 1960s, it is difficult to construct codes that correct classical multiple deletion errors.
In particular, it is tough to construct classical multi deletion error-correcting codes with high code rates.
Sima was the first to achieve the construction of classical 2-deletion codes with optimally low redundancy (parity part),
in 2018 \cite{sima2018twodel,sima2019two}.
Later, codes for three or more deletions with optimally low redundancy were also constructed, also by Sima, in 2019
\cite{sima2019opt,sima2020optimal}.

In the previous researches on quantum communication \cite{chuang1997bosonic,hwang2003quantum,wang2014linear}, the loss of photons is reported.
The loss was modeled as quantum erasure errors and quantum amplitude damping errors, and quantum erasure codes are used for error-correction \cite{bergmann2016quantum,bergmann2016quantumNOON}.
Quantum deletion error is also a model of errors caused by the loss of photons on quantum communication.
Compared to erasure errors,
the deletion model does not assume the knowledge of error indices.
Quantum deletion error is also a partial model of intercept-and-resend.
The intercept-and-resend appears as an attack model for quantum key distribution protocol \cite{bechmann2006eavesdropping,bennett1992experimental,gisin2002quantum,huttner1994information}.
Suppose an analogous theory of classical deletion errors is developed.
In that case, we can expect that quantum deletion error-correcting codes will be able to correct unitary errors as well by considering general unitary errors as intercept-and-resend.
This is discussed at the last paragraph of Section \ref{Deletion}.

The reader may consider utilizing known classical deletion error-correcting codes to construct quantum deletion error-correcting codes.
For example, there is a way to apply VT codes or Sima's codes to CSS code construction.
However, this does not work.
This is because CSS codes require linear codes, but both VT codes and Sima's codes are nonlinear \cite{calderbank1996good,steane1996multiple}.
In general, deletion codes are nonlinear codes.
Therefore, the study of quantum deletion codes requires to create a framework from scratch that is different from existing unitary error-correcting codes.
The author intends not to make the theory far removed from current classical deletion coding theory.
It is rather to be conscious of building a theory that uses current deletion error-correction coding theory.

In the CSS code construction, two classical linear codes with inclusion relations are utilized for unitary error-correction.
Let $C$ and $D$ be classical linear bit-flip error-correcting codes, and let $C$ contain $D$.
In the method of CSS codes, we construct a partition of $C$ by $D$ and construct the basis as a superposition state by elements of the partition.
In this study, we focus on this point, i.e., the partition of $C$.
We construct a partition of the classical deletion error-correcting code that is not assumed to be linear and construct the basis of the quantum code as the superposition state from the elements of the partition.
The new concepts of BRS stable and homogeneous are introduced.
We show that a quantum deletion error-correcting code is constructed from a homogeneous partition.
Furthermore, the encoding and the decoding algorithm are obtained by constructing the homogeneous partition.
Section \ref{sec:HRCconst} constructs classical deletion codes that are ingredients of homogeneous partitions.
As a remark, no partition satisfying BRS stable or homogeneous was obtained from either VT codes or Sima's codes.
The quantum codes constructed in this paper may have a high code rate.
In particular, for any $0 < R < 1$, a quantum single deletion error-correcting code can be constructed with a coding rate greater than or equal to $R$
(Theorem \ref{thm:highRateExistence}).

This manuscript omits the proofs of the statements in the paper.
\section{Preliminaries}\label{Pre}
The paper assumes the fundamental knowledge of quantum information theory and classical coding theory, particularly described in Sections \ref{ssec:FQIT} and \ref{ssec:FCDC}.
This section reviews the definition and notation of technical terms for reading the paper.

\subsection{Fundamentals of Quantum Information Theory}\label{ssec:FQIT}
Let $n$ be an integer greater than or equal to 2 and
$[n]:=\{1,2,\ldots ,n\}$.
For a square matrix $ \rho $ over a complex field $\mathbb{C}$,  $\mathrm{Tr}( \rho )$ denotes
the sum of the diagonal elements of $ \rho $. 

Set $\ket{0},\ket{1}\in \mathbb{C}^2$ as $\ket{0}:=(1,0)^T,\ket{1}:=(0,1)^T$, and 
$\ket{\bm{x}}$ as $\ket{\bm{x}}:=\ket{x_1}\otimes \ket{x_2}\otimes \cdots \otimes \ket{x_n} \in \mathbb{C}^{2 \otimes n}$
for a bit sequence $\bm{x}=x_1x_2\cdots x_n\in \{0,1\}^n$. 
Here $\otimes$ is the tensor product operation, $T$ is the transpose operation, and 
$\mathbb{C}^{2 \otimes n}$ is the $n$th tensor product of $\mathbb{C}^2$, i.e.,
$\mathbb{C}^{2 \otimes n} := (\mathbb{C}^2)^{\otimes n}$.

We denote by $S(\mathbb{C}^{2\otimes n})$ the set of all density matrices of order $2^n$.
A density matrix is employed to represent a quantum state.
The quantum state $\rho$ that relates to $n$-particles/photons $p_1, p_2, \dots, p_n$ is represented in an element of
$S(\mathbb{C}^{2\otimes n})$.
Any state $\rho$ is represented in the following form:
$$\rho=\sum_{\bm{x},\bm{y}\in \{0,1\}^n} \rho_{\bm{x},\bm{y}}\ket{x_1}\bra{y_1}\otimes \cdots \otimes \ket{x_n}\bra{y_n},$$
where $\rho_{\bm{x},\bm{y}} \in \mathbb{C}$, $\bra{0} = (1, 0)$, and $\bra{1} = (0,1)$.
The quantum state of a subsystem that relates to $(n-1)$-particles/photons $p_1, p_2, \dots, p_{i-1}, p_{i+1}, \dots, p_n$ is described by the partial trace defined below.

\begin{Definition}[Partial Trace, $\mathrm{Tr}_i$]
Let $i\in [n]$.
Define a function $\mathrm{Tr}_i:S(\mathbb{C}^{2\otimes n})\rightarrow S(\mathbb{C}^{2\otimes (n-1)})$
as
\begin{align*}
\mathrm{Tr}_i( \rho ):=&\sum_{\bm{x},\bm{y}\in \{0,1\}^n}
\rho_{\bm{x},\bm{y}}\cdot \mathrm{Tr}(\ket{x_i}\bra{y_i}) \ket{x_1}\bra{y_1}\otimes \\
&\cdots \otimes \ket{x_{i-1}}\bra{y_{i-1}}\otimes \ket{x_{i+1}}\bra{y_{i+1}}\otimes \\
&\cdots \otimes \ket{x_n}\bra{y_n},
\end{align*}
where
$$\rho=\sum_{\bm{x},\bm{y}\in \{0,1\}^n} \rho_{\bm{x},\bm{y}}\ket{x_1}\bra{y_1}\otimes \cdots \otimes \ket{x_n}\bra{y_n}$$
and $\mathrm{Tr}( \ket{ x_i } \bra{ y_i } )$ is $1$ if $x_i = y_i$ or $0$ otherwise.
The map $\mathrm{Tr}_i$ is called the partial trace.
\end{Definition}

\begin{Definition}[Projective Measurement, $\mathbb{P}$]
A set $\mathbb{P}:=\{P_0,\ldots, P_{p-1}\}$ of complex square matrices 
$P_0,\ldots, P_{p-1}$ of order $2^n$ is called a projective measurement if
every $P_i$ is a projection matrix and
$$\sum_{0 \le i < p} P_i=\mathbb{I},$$
where $\mathbb{I}$ is the identity matrix of order $2^n$.

If we perform the projective measurement $\mathbb{P}$ on a quantum state $\rho\in S(\mathbb{C}^{2 \otimes n})$,
the probability to obtain an outcome $0 \le k < p$ is $\mathrm{Tr}(P_k \rho)$,
and the state associated with $k$ after the measurement $\rho '$ is 
$$\rho ':=\frac{P_k \rho P_k }{\mathrm{Tr}(P_k \rho)}.$$
\end{Definition}

\begin{Definition}[Code Rate]
Let $Q$ be a quantum code that is a set of pure states, i.e.,
$Q \subset \mathbb{C}^{2 \otimes n}$.
If $Q$ is of dimension $M$ as a complex vector space,
the code rate of $Q$ is defined as $(\log_2 M) / n$. 
\end{Definition}

\subsection{Fundamentals of Classical Deletion Codes}\label{ssec:FCDC}
In coding theory,
a classical single deletion error is described as a map from a bit sequence
$x_1\dots x_{i-1}x_ix_{i+1}\ldots x_n\in \{0,1\}^n$ to a shorter sequence
$x_1\ldots x_{i-1}x_{i+1}\ldots x_n\in \{0,1\}^{n-1}$ for some $i$.
The error shrinks the sequence length.
For example, a single deletion may change a sequence $111000$ to $11000$ or $11100$
and may change another sequence $000000$ to $00000$.
The receiver is not informed the index of error positions.

A similar error is erasure.
The erasure error changes a part of a sequence to the erasure symbols $?$.
The error does not change the sequence length.
For example, a single deletion may change a sequence $111000$ to 
$?11000$, $1?1000$, $11?000$, $111?00$, $1110?0$, or $11100?$.
Hence the receiver can know the index of error positions.

\begin{Definition}[Run]
A run in a sequence $x_1 x_2 \dots x_n$ is 
a maximal subword $x_{i} x_{i+1} \dots x_{j}$ consisting of identical symbols.
Hence, $x_{i-1} \neq x_{i} = x_{i+1} = \dots = x_j \neq x_{j+1}$.
\end{Definition}
For example, the runs of $\bm{x} := 00111011$ are 
$x_1 x_2 = 00$, $x_3 x_4 x_5 = 111$, $x_6 = 0$, and $x_7 x_8 = 11$.

\begin{Definition}[Deletion Surface for Single Deletions, $\delta$]
For an $n$-bit sequence $\bm{a}$,
set $\delta(\bm{a})$ as
the set of $(n-1)$-bit sequences
obtained by classical single deletions to $\bm{a}$.
The set $\delta(\bm{a})$ is called a deletion surface (or a deletion ball)
with its center $\bm{a}$ of radius $1$.
\end{Definition}

\begin{Definition}[Classical Single Deletion Error-Correcting Code]
Let $C$ be a set of bit sequences of the fixed length.
The set $C$ is called a classical single deletion error-correcting code
if $\delta(\bm{x}) \cap \delta(\bm{y}) = \emptyset$ 
for any distinct $\bm{x}, \bm{y} \in C$.
\end{Definition}

\begin{Definition}[Code Rate]
The code rate of $C$ is defined as 
$(\log_2 |C|) / n$,
where $|C|$ is the cardinality of $C$.
\end{Definition}

\begin{Fact}[\cite{levenshtein1966binary} VT Code]
Let $n$ be a positive integer and $a$ an integer.
The following set $\mathrm{VT}_n (a)$ is called a VT code:
$$
\mathrm{VT}_n (a) 
:= \{ x_1 x_2 \dots x_n \in \{ 0, 1\}^n
  \mid x_1 + 2 x_2 + \dots + n x_n \equiv a \pmod{n+1}
   \}.
$$
Remark that each $x_i$ is an integer but not an element of the binary field.
Hence if $x_1 = x_2 = 1$, $x_1 + 2 x_2 = 1 + 2 = 3$.

Levenshtein showed that $\mathrm{VT}_n (a)$ is a classical single deletion error-correcting code for any $n$ and $a$.

For the case $a=0$, the cardinality of $\mathrm{VT}_n (0)$ is greater than or equal to $2^{n} /(n+1)$
\cite{stanley1972study}.
Therefore, the code rate is greater than or equal to
$$
\frac{ n - \log_2 (n+1) }{n} = 1 - \frac{ \log_2 (n+1)}{n}.
$$
It implies that,
for any $0 < R < 1$,
there exists a VT code whose code rate is greater than $R$.
\end{Fact}

An insertion error is a converse operation of a deletion error.
A single insertion error changes
a bit sequence $x_1 x_2 \dots x_n \in \{ 0, 1 \}^n$ to
$x_1 x_2 \dots x_i b x_{i+1} \dots x_n  \in \{ 0, 1 \}^{n+1}$
for some bit $b$.

As Hamming distance is useful for bit-flip error-correction, 
Levenshtein distance is useful for deletion error-correction.

\begin{Definition}[Levenshtein Distance, $d_L$]
Levenshtein distance $d_L(\bm{a}, \bm{b})$ is 
the minimum number of single-character edits, i.e., insertions/deletions, required to change the sequence $\bm{a}$ to 
the other sequence $\bm{b}$.
\end{Definition}

This distance is also known as edit distance.
The following are examples of properties of Levenshtein distance.

\begin{Fact}\label{fact:levendis}\label{fact:levDelEC}
Let $\bm{x}$ and $\bm{y}$ be sequences over a finite alphabet.
\begin{itemize}
\item
$d_{L}( \bm{x}, \bm{y} ) = | \bm{x} | + | \bm{y} | - 2 |\mathrm{lcs}(\bm{x}, \bm{y})|,$
where $| \bm{x} |$ is the length of $\bm{x}$
and $| \mathrm{lcs}(\bm{x}, \bm{y})|$ is the length of the longest common sequences for $\bm{x}$ and $\bm{y}$.
\item The distance $d_L( \bm{x}, \bm{y} )$ is even, if $| \bm{x} | = | \bm{y} |$.
\item Under the assumption $|\bm{x}| = |\bm{y}|$,
$d_L (\bm{x}, \bm{y}) \le 2$ holds
if and only if 
$\delta( \bm{x}) \cap \delta( \bm{y}) \neq \emptyset$
holds.
\item Under the assumption $|\bm{x}| = |\bm{y}|$,
 $d_L (\bm{x}, \bm{y}) \ge 4$ holds
if and only if 
$\delta( \bm{x}) \cap \delta( \bm{y}) = \emptyset$
holds.
\item For a set $C$ of fixed length sequences,
$C$ is a classical single deletion error-correcting code
if and only if
$d_L (\bm{a}, \bm{b}) \ge 4$ holds
for any distinct $\bm{a}, \bm{b} \in C$.
Note that the minimum distance decoding for Levenshtein distance is an example of its decoder.
\end{itemize}
\end{Fact}

For a positive integer $t$,
a $t$-deletion error is a combination of $t$ single-deletion errors.
The error changes an $n$-bit sequence to an $(n-t)$-bit sequence.
Similarly, 
a $t$-insertion error is a combination of $t$ single-insertion errors.
The error changes an $n$-bit sequence to an $(n+t)$-bit sequence.

The following mentions the importance of study on deletion error-correcting codes.
\begin{Fact}\label{fact:classicalEquiv}
Let $C$ be a set of bit sequences and $t$ a positive integer.
The following are equivalent.
\begin{itemize}
\item $C$ is a code that can correct any at most $t$-deletion error.
\item $C$ is a code that can correct any at most $t$-insertion error.
\item $C$ is a code that can correct any combination error of an $s_1$-deletion error
and an $s_2$-insertion error with $s_1 + s_2 \le t$.
\item The minimum Levenshtein distance of $C$ is greater than or equal to $2t + 1$,
i.e., for any distinct $\bm{x}, \bm{y} \in C$, $d_L ( \bm{x}, \bm{y} ) \ge 2t+1$ holds.
\end{itemize}
\end{Fact}

A $t$-bit flip (substitution) errors are implementable by
a combination of $t$-deletion and $t$-insertion errors.
For example,
a bit sequence $x_1 x_2 \dots x_n$ is changed to
$x_1 x_2 \dots x_{i-1} x_{i+1} \dots x_n$ by a single-deletion error
and 
it is changed to 
$x_1 x_2 \dots x_{i-1} b x_{i+1} \dots x_n$ by a single-insertion error,
where $b$ is the bit flipped symbol of $x_i$.
By combining Fact \ref{fact:classicalEquiv} and the flip errors,
research on deletion error-correction is directly applicable
to research on all deletion, insertion, and bit-flip error-correction.

\section{Quantum Deletion Error}\label{Deletion}
The loss of photons on quantum communication has been reported in various papers \cite{chuang1997bosonic,hwang2003quantum,wang2014linear}.
Quantum deletion error is a model of errors caused by the loss of photons in quantum communication.

This section defines and observes deletion errors for a quantum state.
The previous works, e.g., \cite{nakayama2020first,ouyang2021permutation,shibayama2020new}, identify deletion error with partial trace.
Here, more explanation between them is provided.

\begin{Definition}[Quantum Single Deletion Error, $D_i$] \label{deletionError}
A quantum single deletion error is a map $D_i$
that changes a quantum state
in $S(\mathbb{C}^{2\otimes n})$ of
$n$-ordered particles $p_1, p_2, \dots, p_n$
to a quantum state in $S(\mathbb{C}^{2 \otimes (n-1)})$
of
$(n-1)$-ordered particles $p_1, p_2, \dots, p_{i-1}, p_{i+1}, \dots, p_n$,
i.e.,
$$
D_i (p_1, p_2, \dots, p_n) = p_1, p_2, \dots, p_{i-1}, p_{i+1}, \dots, p_n.
$$
\end{Definition}
By denoting the state of $p_1, p_2, \dots, p_n$ by $\rho$
and the state of $D_i (p_1, p_2, \dots, p_n)$ by $\rho'$,
then we have
$$
\mathrm{Tr}_i ( \rho ) = \rho'.
$$

\begin{Remark}\label{rem:deletionError}
The state after quantum single deletion error is described by a partial trace.
The quantum single deletion corresponding to $p_i$
is described by $\mathrm{Tr}_i$.
In this sense, 
it may be preferable to denote
the operation that represents a quantum single deletion by $D_i$,
instead of $\mathrm{Tr}_i$.
For example, for the state $\rho$ of $p_1, p_2, \dots, p_n$,
we allow to say
$$
D_i (\rho) = \mathrm{Tr}_i (\rho).
$$
\end{Remark}

A quantum single insertion error is also defined in a similar way to 
a classical insertion error.
It is an error that changes 
$n$-ordered particles $p_1, p_2, \dots, p_n$
to 
$(n+1)$-ordered particles $p_1, p_2, \dots, p_{i}, q, p_{i+1}, \dots,  p_n$
for some $q$.

The author conjectures that there is a quantum version of 
Fact \ref{fact:classicalEquiv}.

In other words,
if a quantum code can correct any at most $t$-deletion errors,
the code can correct any combinations of 
quantum $s_1$-deletions and $s_2$-insertions, where $s_1 + s_2 \le t$.

The author remarks that 
any single unitary error is a combination
of a quantum single deletion and single insertion errors.
Let $\rho$ be the state of $n$-ordered particles $p_1, p_2, \dots, p_n$
and see a kind of intercept-and-resend operations to $\rho$.
A quantum single deletion $D_i$ 
changes
$n$-ordered particles $p_1, p_2, \dots, p_n$
to
$(n-1)$-ordered particles $p_1, p_2, \dots, p_{i-1}, p_{i+1}, \dots, p_n$
with the missing particle $p_i$.
Then consider performing a unitary operator $U$ to the particle $p_i$.
After the unitary operation, insert $p_i$ back to the original $i$th position.
Then $(n-1)$-ordered particles $p_1, p_2, \dots, p_{i-1}, p_{i+1}, \dots, p_n$
are changed to
$n$ ordered particles $p_1, p_2, \dots, p_n$.
Readers may recall the intercept-and-resend attack from these actions.
In other words, it is an attack where one extracts and
performs a unitary transformation $U$ to $p_i$, and then returns $p_i$ into the $n-1$ particles.
Then the state of $n$-ordered particles $p_1, p_2, \dots, p_n$
is $(I \otimes \dots \otimes I \otimes U \otimes I \otimes \dots \otimes I) (\rho)$,
where $I$ is the identity matrix.
It is a single unitary error.

\section{Generalization from Single Qubit Messages to Multi Qubit Messages}\label{Construct}
The first quantum single deletion error-correcting code was constructed by \cite{nakayama2020first}.
The code length is $8$ and its dimension is $2$.
Its code rate is $1/8$.
The next code was constructed by \cite{ManabuHagiwara20212020XBL0191}.
It is shown that this code is the theoretically shortest length quantum deletion error-correcting code.
The code length is $4$ and its dimension is $2$.
Hence the code rate is $1/4$.
After these two codes, 
Shibayama \cite{shibayama2020new},
Shibayama and Hagiwara \cite{shibayama2021permutation},
Ouyang \cite{ouyang2021permutation},
and
Matsumoto and Hagiwara \cite{matsumoto2022constructions}
proposed classes of quantum deletion error-correcting codes.
There have not been quantum deletion error-correcting codes with a high code rate.
Furthermore, for a long length,
all the code rates were closed to $0$.
This paper provides a new class of quantum single deletion error-correcting codes
that contains high rate codes.
In particular, for any $0 < R < 1$,
there exists a code whose code rate is greater than $R$.

In this section, we define an encoder (see \ref{ss:encoder}) and a decoder (see \ref{ss:decoder}, \ref{ss:proj}) for quantum states
by generalizing three conditions, say \cratio, \coutd, and \cind (see \ref{ss:threeCond}).
Afterward, we discuss the error-correctability
by the encoder and the decoder under the three conditions (see \ref{subsec:ecp}).

The three conditions were originally defined by \cite{nakayama2020single}.
The original conditions are only applicable for constructing quantum codes of dimension $2$,
i.e., the message is a single qubit.
This generalization allows us to encode a multi-qubits message.

Throughout of this section,
$\mathcal{X}$ denotes a family set of bit sequences,
i.e.,
$\mathcal{X}$ is a set of subsets of $\{ 0, 1 \}^n$ for some $n$.
It is assumed that the empty set $\emptyset$ does not belong to $\mathcal{X}$,
and $\mathcal{X}$ is not the empty set,
i.e., $\emptyset \notin \mathcal{X}$
and $| \mathcal{X} | > 0$.

\subsection{Encoder}\label{ss:encoder}
At first, our encoder is obtained by a family set.
The encoder is a generalization of previously defined encoders
for deletion error-correction,
e.g., \cite{nakayama2020first,shibayama2020new,shibayama2021permutation,hagiwara2020four}.
It is also a generalization of encoders for CSS codes
\cite{calderbank1996good,steane1996multiple}.

\begin{Definition}[Encoder, $\mathrm{Enc}_{\mathcal{X}}$]
Let $\mathcal{X} = \{ X^{(m)} \}_{ 0 \le m < M } $ be a family set of $\{0,1\}^n$ 
such that 
$X^{(i)} \cap X^{(j)} =\emptyset$ for $i \neq j$ and $M \ge 2$.

Define an encoder $\mathrm{Enc}_{\mathcal{X}} : 
S(\mathbb{C}^{2 \otimes \lceil \log_2 M \rceil} ) \rightarrow S(\mathbb{C}^{2 \otimes n})$
as 
$$\mathrm{Enc}_{\mathcal{X}}(\sigma) :=
\ket{\Psi}\bra{\Psi},$$
where 
$\lceil \log_2 M \rceil$ is the smallest integer greater than or equal to $\log_2 M$,
$\sigma:=\ket{\psi} \bra{\psi} \in S(\mathbb{C}^{2 \otimes \lceil \log_2 M \rceil})$
is a quantum message
with a unit vector 
$$\ket{\psi}:= \sum_{0 \le m < M} \alpha_m \ket{\bm{m}} \in \mathbb{C}^{2 \otimes \lceil \log_2 M \rceil},$$
$\bm{m} \in \{0, 1\}^{\lceil \log_2 M \rceil} $ is the binary expression of $m$ in 
$\lceil \log_2 M \rceil$-bits,
and
$$\ket{\Psi}
:=
\sum_{0 \le m < M}
\alpha_m
\left( 
\frac{1}{\sqrt{|X^{(m)}|}} 
\sum_{ \bm{x} \in X^{(m)}} \ket{\bm{x}} 
\right).
$$

Hence a pure state $| \bm{m} \rangle$ is encoded to
$$\frac{1}{\sqrt{|X^{(m)}|}} \sum_{ \bm{x} \in X^{(m)}} \ket{\bm{x}} \in \mathbb{C}^{2 \otimes n}.$$

Since $X^{(i)} \cap X^{(j)} = \emptyset$ for $i \neq j$,
the encoded states of $| \bm{i} \rangle$ and $| \bm{j} \rangle$
are orthogonal.
\end{Definition}

The partition is a terminology in basic set theory.
\begin{Definition}[Partition]
Let $C$ be a set and
$X^{(m)}$ ($0 \le m  <M$) a non-empty subset of $C$ for some integer $M$.

The family set $\{ X^{(m)} \mid 0 \le m < M \}$ is called a partition of $C$
if
\begin{itemize}
\item $C = \bigcup_{0 \le m < M} X^{(m)}$,
\item $X^{(m)} \cap X^{(m')} = \emptyset$ for $m \neq m'$.
\end{itemize}
\end{Definition}

Remark that if a family set $\mathcal{X}$ is a partition of some set,
the encoder $\mathrm{Enc}_{\mathcal{X}}$
keeps the inner product of quantum states
before and after the encoding.

\begin{Example}[Shortest Single Deletion Code \cite{hagiwara2020four}]
Set $X^{(0)} := \{ 0000, 1111 \}$,
 $X^{(1)} := \{ 0011, 0101, 0110, 1001, 1010, 1100 \}$
 and $\mathcal{X} := \{ X^{(0)}, X^{(1)} \}$.
The encoder $\mathrm{Enc}_{\mathcal{X} }$
is same as the encoder of the shortest quantum single deletion code.
\end{Example}

\begin{Example}[Quantum Hamming Code \cite{steane1996multiple}]
The quantum Hamming code is an example of CSS codes.
Let $C$ be a classical Hamming code
and $D$ be the set of even Hamming weight codewords in $C$.

Set
$X^{(0)} := D$, $X^{(1)} := C \setminus D$ and $\mathcal{X}:= \{ X^{(0)}, X^{(1)} \}$.
Hence
$$
X^{(0)} 
=
\{ 0000000, 0001111, 0111100, 0110011, 1010101, 1011010, 1100110, 1101001 \},
$$
and
$$
X^{(1)} 
=
\{ 1111111, 1110000, 1000011, 1001100, 0101010, 0100101, 0011001, 0010110 \}.
$$
Then $\mathrm{Enc}_{\mathcal{X}}$ is an encoder for quantum Hamming code.

Note that $\mathcal{X}$ is a partition of the classical Hamming code $C$
and $| X^{(1)} | = | X^{(2)} |$.
\end{Example}

\subsection{The Three Conditions for General Dimension}\label{ss:threeCond}
Next, a generalization \cratio, \coutd, and \cind  of the three conditions \cite{nakayama2020single} are defined
for constructing quantum codes of general dimension $M$.
This generalization is equivalent to the original conditions if $M=2$.
For the preparation, the symbols $\Delta_{i,b}$ and $X_{I, b}$ are given as follows.
\begin{Definition}[$\Delta_{i,b}$, $X_{I, b}$]
Let $i\in [n]$, $b \in \{0,1\}$
and $I \subset [n]$ with $I \neq \emptyset, [n]$.

For any $X \subset \{0,1\}^n$,
define the set $\Delta_{i,b}(X) \subset \{0,1\}^{n-1}$ as
\begin{align*}
\Delta_{i,b}(X):=\{(a_1,\ldots ,a_{i-1},a_{i+1},\ldots ,a_n)\in \{0,1\}^{n-1}|\\
(a_1,\ldots ,a_{i-1},b,a_{i+1},\ldots ,a_n) \in X\}.
\end{align*}
We call the set $\Delta_{i,b}(X)$ the $(i,b)$-deletion set of $X$.

For $X \subset \{0,1\}^n$,
$X_{I, b}$ is defined as
 $$X_{I, b} := 
 \bigcap_{i \in I} \Delta_{i,b}(X) 
  \cap
 \bigcap_{i \in I^c} \Delta_{i,b}(X)^c,$$
where $^c$ is the complement operator,
 in particular, $I^c = [n] \setminus I$
 and $\Delta_{i, b}(X)^c = \{0,1\}^{n-1} \setminus \Delta_{i,b}(X)$.
For $I = [n]$, we additionally define
$$X_{[n], b} := \bigcap_{i \in [n]} \Delta_{i,b}(X) .$$
\end{Definition}

The following holds from 
the definition of $\delta$ and $\Delta_{i,b}$:
$$ \delta(\bm{a})= \bigcup_{i \in [n], b=0,1 } \Delta_{i, b}( \{ \bm{a} \} ),$$
for an $n$-bit sequence $\bm{a}$.

\begin{Example}
Set $X := \{ 0101, 1010, 0100, 1111 \}$.
$\Delta_{i, b}(X)$ is the set of sequences
that is obtained by deletion at the $i$th position if the $i$th entry is $b$.
Therefore
\begin{align*}
\Delta_{1,0}(X) &= \{101, 100 \}, & \Delta_{1,1}(X) &= \{010, 111 \},\\
\Delta_{2,0}(X) &= \{110 \},      & \Delta_{2,1}(X) &= \{001, 000, 111 \},\\
\Delta_{3,0}(X) &= \{011, 010 \}, & \Delta_{3,1}(X) &= \{100, 111 \},\\
\Delta_{4,0}(X) &= \{101, 010 \}, & \Delta_{4,1}(X) &= \{010, 111 \}.
\end{align*}

The union of $\Delta_{i,0}(X)$ consists of five elements:
$$
\bigcup_{1 \le i \le 4} \Delta_{i,0}(X) 
=
\{ 010, 011, 100, 101, 110 \}.
$$
Since $010$ in the union belongs to 
$\Delta_{3,0}(X)$ and $\Delta_{4,0}(X)$,
$$010 \in X_{ \{3,4\}, 0 }.$$
Similarly,
\begin{align*}
011 & \in X_{ \{3\}, 0}, & 100 & \in X_{ \{1 \}, 0},\\
101 & \in X_{ \{1,4\}, 0}, & 110 & \in X_{ \{2 \}, 0}.
\end{align*}
\end{Example}

\begin{Definition}[Conditions \cratio, \coutd, and \cind]
For a family set $\mathcal{X} = \{ X^{(m)} \}_{0 \le m < M} $ of $\{ 0,1 \}^n$,
define three conditions \cratio, \coutd, and \cind as follows.
\begin{itemize}
\item[]\cratio: Ratio Condition and the ratio $\lambda_{I,b}$:\\
For any $X^{(m_1)}, X^{(m_2)} \in \mathcal{X}$,
the following holds:
$$
|X^{(m_1)}||X^{(m_2)}_{I, b}|=|X^{(m_2)}||X^{(m_1)}_{I, b}|,
$$
for any non-empty set $I \subset [n]$ and any $b \in \{0,1\}$.

Since the ratio $| X^{(m)}_{I, b} | / |X^{(m)}|$ is independent in the choice of $0 \le m < M$,
we denote the ratio by $\lambda_{I, b}$.

\item[]\coutd: External Distance Condition:\\
For any distinct $0 \le m_1, m_2 < M$,
the following holds:
$$
| \Delta_{i_1, b_1}( X^{(m_1)} ) \cap \Delta_{i_2, b_2}( X^{(m_2)} )| = 0,
$$
for any $i_1, i_2 \in [n]$ and any $b_1, b_2 \in \{0,1\}$.

\item[]\cind: Internal Distance Condition:\\
For any $X^{(m)} \in \mathcal{X}$,
the following holds:
$$
| \Delta_{i_1, 0}( X^{(m)} ) \cap \Delta_{i_2, 1}( X^{(m)} )| = 0,
$$
for any $i_1, i_2 \in [n]$.
\end{itemize}

We call $\mathcal{X}$ a (Ci)$^{-}$ family set for $i=1,2,3$ if
$\mathcal{X}$ satisfies the condition (Ci)$^{-}$
and $\mathcal{X}$ is not the empty set.
\end{Definition}

The condition \cratio is mainly corresponding to the state after
measurement in decoding (see \ref{subsec:ecp})
and the conditions \coutd and \cind are mainly corresponding to
define projective measurement (see \ref{ss:proj} and \ref{ss:decoder}).

\begin{Lemma}\label{delIncludion}
Let $X$ be a non-empty subset of $\{0, 1\}^n$,
$I \subset [n]$,
and
$b \in \{0, 1\}$.

$X_{I, b} \cap \Delta_{i,b}(X)$ is equal to $X_{I, b}$ if $i \in I$.
In other words, $X_{I, b} \subset \Delta_{i, b}( X )$ if $i \in I$.
On the other hand, $X_{I, b} \cap \Delta_{i,b}(X) = \emptyset$ if $i \notin I$.
\end{Lemma}

\begin{Lemma}\label{lemma:Jofx}
Let 
$X$ be a non-empty subset of $\{0, 1\}^n$,
$1 \le j \le [n]$ and $b \in \{0, 1\}$.

For $\bm{x} \in \Delta_{j, b}(X)$,
$\bm{x} \in X_{J, b}$
if and only if
$J = \{ i \mid \bm{x} \in \Delta_{i, b}(X) \}$.
Hence $J$ is unique for the fixed $j$ and $b$.
\end{Lemma}

\begin{Lemma}\label{lemma:delEqXIb}
Let $n$ be a positive integer,
$X$ a set of binary sequences of length $n$,
$b$ a bit,
and $j \in [n]$.

$$ \Delta_{j, b}(X) = \bigcup_{I \subset [n], j \in I} X_{I, b}.$$

Furthermore, for fixed $j$ and $b$,
$\{ X_{I, b} \mid j \in I \}$ is a partition of $\Delta_{j, b}$.
\end{Lemma}

\subsection{Projective Measurement for Decoding}\label{ss:proj}
This section defines the projective measurement from
a \cratio, \coutd, and \cind family set $\mathcal{X}$.
The projection measurement is used as a part of decoding.

\begin{Definition}[$\mathbb{P}'_{\mathcal{X}}$ and Projective Measurement, $\mathbb{P}_{\mathcal{X}}$]
Let $\mathcal{X} = \{ X^{(m)} \}_{0 \le m < M}$ be 
a \coutd and \cind family set.

Then define a set $\mathbb{P}_{\mathcal{X}}'$ of matrices as follows.
\begin{equation*}
\begin{split}
\mathbb{P}_{\mathcal{X}}'
:=\{ P_{I,b} \mid I\subset [n], I \neq \emptyset, b \in \{0,1\}, P_{I,b} \neq O \},
\end{split}
\end{equation*}
where 
$$P_{I,b}:=\sum_{0 \le m < M} \sum_{\bm{x} \in X^{(m)}_{I, b} }\ket{\bm x}\bra{\bm x},$$
and $O$ is the zero matrix of degree $2^{n-1}$.

Define $\mathbb{P}_{\mathcal{X}}$ as follows.
\begin{equation*}
\mathbb{P}_{\mathcal{X}}:=\mathbb{P}_{\mathcal{X}}' \cup \{P_{\emptyset}
 := \mathbb{I}-\sum_{P \in \mathbb{P}_{\mathcal{X}}'}P\},
\end{equation*}
where $\mathbb{I}$ is the identity matrix of order $2^{n-1}$.
\end{Definition}

From here, we prove that $\mathbb{P}_{\mathcal{X}}$ is a projective measurement.

\begin{Lemma}\label{projection2}
Let $\{ X^{(m)} \}_{0 \le m < M}$ be a \coutd family set.

Then for any non-empty sets $I_1, I_2 \subset [n]$ and any bits $b_1, b_2\in \{0,1\}$,
$$X^{(m_1)}_{I_1,b_1}\cap X^{(m_2)}_{I_2,b_2}=\emptyset,$$
if $m_1 \neq m_2$.
In particular, $\braket{ \bm{x} | \tilde{\bm{x}} } = 0$
for $\bm{x} \in X_{I_1, b_1}^{(m_1)}$, and $\tilde{\bm{x}} \in X_{I_2, b_2}^{(m_2)}.$
\end{Lemma}

\begin{Lemma}\label{projection}
Let $\{ X^{(m)} \}_{0 \le m < M} $ is a \cind family set.

Then for any distinct $(I_1, b_1)$ and $(I_2, b_2)$ with non-empty sets $I_1, I_2 \subset [n]$,
$$X^{(m)}_{I_1, b_1}\cap X^{(m)}_{I_2, b_2}=\emptyset,$$
for all $0 \le m < M$.
In particular, $\braket{ \bm{x} | \tilde{\bm{x}} } = 0$
for $\bm{x} \in X_{I_1, b_1}^{(m)}$, and $\tilde{\bm{x}} \in X_{I_2, b_2}^{(m)}.$
\end{Lemma}

\begin{Lemma}\label{lemma:Pproj}
Let $\mathcal{X} = \{ X^{(m)} \}_{0 \le m < M}$ be 
a \coutd and \cind family set.

For any $P \in \mathbb{P}_{\mathcal{X}}'$,
$P^\dagger = P$ and $P^2 = P$.
In other words, any element of $\mathbb{P}_{\mathcal{X}}'$ is a projection matrix.
\end{Lemma}

\begin{Lemma}\label{lemma:ptildepZero}
Let $\mathcal{X}$ be a \coutd and \cind family set
and $P, \tilde{P} \in \mathbb{P}_{\mathcal{X}}'$.

If $P \neq \tilde{P}$,
then $P \tilde{P} = O$.
\end{Lemma}

\begin{Lemma}\label{orthogonal}
Let $\mathcal{X} = \{ X^{(m)} \}_{0 \le m < M}$ be 
a \coutd and \cind family set.

Then the following holds:
$$
\sum_{P\in \mathbb{P}_{\mathcal{X}}'} \sum_{\tilde{P}\in \mathbb{P}_{\mathcal{X}}'} P \tilde{P}
=\sum_{P\in \mathbb{P}_{\mathcal{X}}'}P.
$$
\end{Lemma}

\begin{Theorem}\label{thm:projM}
Let $\mathcal{X}$ be a \coutd and \cind family set.

Then the set $\mathbb{P}_{\mathcal{X}}$ is the projective measurement.
\end{Theorem}

\subsection{Decoder}\label{ss:decoder}
Our decoder is a combination of projective measurement and recovery operation.
The projective measurement has been already obtained in the previous subsection by Theorem \ref{thm:projM}.
The error-correcting operation is introduced as below.

Throughout this section, $I$ denotes a non-empty subset of $[n]$.

\begin{Definition}[$\ket{ \psi^{(m)}_{I, b} }$]
Let $\mathcal{X}$ be a \cratio, \coutd, \cind family set,
$b \in \{0, 1\}$, and $\lambda_{I, b} \neq 0$.
Set the state $\ket{ \psi^{(m)}_{I, b} }$ as
$$
\ket{ \psi^{(m)}_{I, b} }
:= 
 \frac{ 1 } 
      {  \sqrt{| X^{(m)}_{I, b} |}   }
      \sum_{\bm{x} \in X^{(m)}_{I,b} } \ket{\bm x}
 \in \mathbb{C}^{2 \otimes (n-1)}.
$$
\end{Definition}

\begin{Lemma}\label{lemma:psiOrtho}
For fixed $I$ and $b \in \{0,1\}$,
$$
\braket{ \psi^{(m)}_{I, b} \mid \psi^{(m')}_{I, b}}
= 
\begin{cases}
1 &  m=m',\\
0 &  \text{otherwise}.
\end{cases}
$$ 
\end{Lemma}

\begin{Definition}[Recovery Operator, $U_{I,b}$] \label{errorCorrectingOperator}
Let $\mathcal{X}$ be a \cratio, \coutd, and \cind family set.
Let $I \subset [n]$ and $b \in \{0, 1\}$
such that $\lambda_{I, b} \neq 0$.

By Lemma \ref{lemma:psiOrtho},
there exists a unitary matrix $U_{I,b}$ such that
$$
U_{I,b} \ket{ \psi^{(m)}_{I, b} }
 = \ket{\bm{0} \bm{m}},
$$
for all $0 \le m < M$.
Here $\bm{m}$ is the binary expression of $0 \le m < M$ in $\lceil \log_2 M \rceil$bits,
and $\bm{0}$ is the zero vector of length $n - 1 - \lceil \log_2 M \rceil$.
We call the matrix $U_{I,b}$ a recovery operator.
\end{Definition}

We clarify the timing when $\lambda_{I, b} \neq 0$ 
in Definition \ref{errorCorrectingOperator}
happens by the statement below.

\begin{Lemma}
Let $\{ X^{(m)} \}_{0 \le m < M} $ be a \cratio family set.
Let $I \subset [n]$ and $b \in \{0, 1\}$.

$\lambda_{I, b} = 0$
if and only if 
$X_{I, b}^{(m)} = \emptyset $ for all $0 \le m < M$.
\end{Lemma}

Finally, the decoder is defined.
The next section \ref{subsec:ecp} justifies
the encoder and the decoder under the three conditions.

\begin{Definition}[Decoder, $\mathrm{Dec}_{\mathcal{X}}$] \label{decoding}
Let $\mathcal{X}$ be a \cratio, \coutd, and \cind family set
and $\mathbb{P}_{\mathcal{X}}$ the projective measurement.
Define a function 
$\mathrm{Dec}_{\mathcal{X}}:S(\mathbb{C}^{2 \otimes (n-1)})\rightarrow S(\mathbb{C}^{2 \otimes \lceil \log_2 M \rceil})$ as
a map that assigns $\rho '\in S(\mathbb{C}^{2\otimes (n-1)})$ to $\sigma '\in S(\mathbb{C}^{2 \otimes \lceil \log_2 M \rceil})$ which
is constructed by the following procedure.

\begin{enumerate}
\item Perform the projective measurement $\mathbb{P}_{\mathcal{X}}$ under the state $\rho '$ (Theorem \ref{thm:projM}).
Assume that the outcome is $(I,b)$ and that the state after the measurement is $\rho '_{I,b}$.
\item Let $\tilde{\rho}:=U_{I,b}\rho '_{I,b} U^\dagger_{I,b}$.
Here $U_{I,b}$ is the recovery operator.
\item At last, 
return $\sigma ':=\underbrace{\mathrm{Tr}_1\circ \cdots \circ \mathrm{Tr}_1}_{n - \lceil \log_2 M \rceil - 1 \ \mathrm{times}}(\tilde{\rho})$.
\end{enumerate}
\end{Definition}

\subsection{Error-Correctability}\label{subsec:ecp}
The encoder, the decoder, and the three conditions are justified by the following statement:
\begin{Theorem}\label{thm:mainGeneralConstruction}
Let $\mathcal{X}$ be a \cratio, \coutd, and \cind family set.
Then for any pure state $\sigma \in S(\mathbb{C}^{2 \otimes \lceil \log_2 M \rceil})$ and 
any deletion position $i \in [n]$,
$$\mathrm{Dec}_{\mathcal{X}} \circ D_i\circ \mathrm{Enc}_{\mathcal{X}}(\sigma)
=\sigma.$$
Here the symbol $\circ$ represents the composition of functions.
In other words, 
the image of $\mathrm{Enc}_{\mathcal{X}}(S(\mathbb{C}^{2 \otimes \lceil \log_2 M \rceil}))$ 
for the pure states
is  a single quantum deletion error-correcting code
with the decoder $\mathrm{Dec}_{\mathcal{X}}$.
\end{Theorem}

We obtain the following probabilistic aspects of $\lambda_{I, b}$.
\begin{Corollary}
For a fixed $i \in [n]$, we have the following.
\begin{itemize}
\item $\lambda_{I, b} \ge 0$.
\item $\sum_{b \in \{0,1\}} \sum_{i \in I \subset [n]} \lambda_{I, b} = 1$.
\end{itemize}
\end{Corollary}

\section{Classical Deletion Error-Correcting Codes, BRS Stable and Homogeneous}
In this section, we provide a sufficient condition for a partition of a classical deletion error-correcting code to satisfy the three conditions.
The sufficient condition is named a ``homogeneous'' partition in this paper.
In preparation, we also introduce the term ``BRS stable.''
An example of a homogeneous partition of a classical single-deletion error-correcting code is given in the next section.

Just as quantum codes for correcting unitary errors were constructed from classical 
bit-flip error-correcting codes, the author develops theorems that a quantum code for correcting quantum deletion errors can be constructed from classical deletion error-correcting codes.

This section assumes that $\mathcal{X}$ denotes a family set of binary sequences of a fixed length,
the empty set does not belongs to $\mathcal{X}$ and $\mathcal{X}$ is non-empty, 
i.e., $\emptyset \notin \mathcal{X}$ and $| \mathcal{X} | > 0$.

As preparation to define ``BRS stable'' and a ``homogeneous partition,'' $b$-run support $\mathrm{R}_b$ is defined here.
\begin{Definition}[$b$-Run Support, $\mathrm{R}_b$]
Let $X$ be a set of binary sequences of length $n$
and let $\bm{x} = x_1 x_2 \dots x_n \in X$.

A consecutive integer set $I = \{j, j+1, \dots, h-1, h \}$ is called
a $b$-run support for $\bm{x}$
if $I$ is the index set of a run of $\bm{x}$
and $b = x_j = \dots =x_h$.
In other words,
\begin{itemize}
\item $x_i = b$ for any $i \in I$,
\item $x_{j-1} \neq b$, if $1 < j$,
\item $x_{h+1} \neq b$, if $h < n$.
\end{itemize}
The set of $b$-run supports for $\bm{x}$ is denoted by $\mathrm{R}_b ( \bm{x} )$.

For a set $X$ of binary sequences,
$\mathrm{R}_b ( X )$ is defined as the following ``multiset,''
$$
\mathrm{R}_b ( X ) := \bigcup_{\bm{x} \in X} \mathrm{R}_b ( \bm{x} ).
$$
\end{Definition}

\begin{Example}
Let $X = \{ 0001, 0011, 0101, 0111 \}$.
The $0$-run supports for the elements of $X$ are
\begin{align*}
\mathrm{R}_0 ( 0001 ) &= \{ \{1,2,3\} \}, \\
\mathrm{R}_0 ( 0011 ) &= \{ \{1,2\} \}, \\
\mathrm{R}_0 ( 0101 ) &= \{ \{1\}, \{3\} \}, \\
\mathrm{R}_0 ( 0111 ) &= \{ \{1\} \}.
\end{align*}
Therefore, the multiset $\mathrm{R}_0 (X)$ is
\begin{align*}
\mathrm{R}_0 (X) 
 &= \{ \{1,2,3\} \} \cup \{ \{1,2\} \} \cup \{ \{1\}, \{3\} \} \cup \{ \{1\} \} \\
 &= \{ \{1\}, \{1\}, \{3\}, \{1,2\}, \{1,2,3\} \}.
\end{align*}
\end{Example}

\begin{Definition}[BRS stable]
We call a family set $\mathcal{X}$ bit run support (BRS) stable
if $R_b (X) = R_b (Y)$ holds for any $X, Y \in \mathcal{X}$
and any bit $b \in \{0, 1\}$.
\end{Definition}

\begin{Example}
Set 
$\mathcal{X} 
 := \{ \{ 000101, 010111 \}, \{ 010101, 000111 \} \}.$

The $0/1$-run supports for $\{ 000101, 010111 \} \in \mathcal{X}$ are
\begin{align*}
\mathrm{R}_0 (\{ 000101, 010111 \}) 
 & = \{ \{1,2,3\}, \{5\}, \{1\}, \{3\} \}, \\
\mathrm{R}_1 (\{ 000101, 010111 \}) 
 & = \{ \{4\}, \{6\}, \{2\}, \{4,5,6\} \}. 
\end{align*}

The $0/1$-bit run supports for $\{ 010101, 000111 \}$ are
\begin{align*}
\mathrm{R}_0 (\{ 010101, 000111 \}) 
 & = \{ \{1\},\{3\}, \{5\}, \{1,2,3\} \}, \\
\mathrm{R}_1 (\{ 010101, 000111 \}) 
 & = \{ \{2\}, \{4\}, \{6\}, \{4,5,6\} \}.
\end{align*}
 
They have the same bit run supports.
Hence $\mathcal{X}$ is BRS stable.
\end{Example}

Here we introduce ``homogeneous.''
\begin{Definition}[Homogeneous]
Let $\mathcal{X}$ be a family set of binary sequences of a fixed length
and $C$ a classical deletion error-correcting code.

$\mathcal{X}$ is called a homogeneous partition of $C$,
if 1. $\mathcal{X}$ is a partition of $C$,
 2. the cardinality of any element in $\mathcal{X}$ is constant,
 i.e.,
  $|X|=|Y|$ for any $X, Y \in \mathcal{X}$,
and
 3. $\mathcal{X}$ is BRS stable.
\end{Definition}

Let us see the relation between classical deletion error-correctability and
the three conditions.

\begin{Theorem}\label{thm:cindSUF}
If any element $X \in \mathcal{X}$ is a classical single deletion 
error-correcting code,
then $\mathcal{X}$ is a \cind family set.
\end{Theorem}

Note that the following statement assumes a weaker condition than the classical single deletion 
error-correctability.
\begin{Theorem}\label{thm:coutdSUF}
Let a family set $\mathcal{X}$ satisfy $X \cap X' = \emptyset$
for any distinct $X, X' \in \mathcal{X}$.

If $d_L( \bm{x}, \bm{y} ) \ge 4$ holds for any $\bm{x} \in X$ and $\bm{y} \in X'$,
then 
$\mathcal{X}$ is a \coutd family set.
\end{Theorem}

\begin{Theorem}\label{thm:cratioSUF}
Let $\mathcal{X}$ be a BRS stable family set such that
 any element $X \in \mathcal{X}$ is a classical single deletion error-correcting code.

If $|X| = |Y|$ holds for any $X, Y \in \mathcal{X}$,
then $\mathcal{X}$ is a \cratio family set.
\end{Theorem}

\begin{Corollary}\label{cor:corOfMain}
A homogeneous partition $\mathcal{X}$ of a classical single deletion error-correcting code
satisfies \cratio, \coutd, and \cind.
\end{Corollary}

\section{High Rate Code Construction}\label{sec:HRCconst}
In the previous section, 
we showed that a homogeneous partition of a classical single deletion code gives a quantum single deletion code.
In this section, we construct an example of a homogeneous partition.

Throughout this section, $t$, $N$ and $E$ denote positive integers
such that $N$ is a multiple of $2^E$.
In Section \ref{sec:ecTodc},
we provides a construction method of a classical $t$-deletion error-correcting code $C$
from $t$-erasure error-correcting code over $2^E$-ary code alphabet.
The code length is $(E+2t)N$.
In Section \ref{sec:InstHRC}, we apply $t :=1$ and $\mathbb{Z}_{2^E}$ for the code alphabet to the construction method,
where $\mathbb{Z}_{2^E}$ is the quotient ring of integers.
Then a partition $\mathcal{X}$ of $C$ is given.
The partition $\mathcal{X}$ consists of $2^{E (N-2)}$ sets.
We prove that $\mathcal{X}$ is a homogeneous partition.
Hence, we obtain a quantum code of length $(E+2)N$ and dimension $2^{E(N-2)}$.
The construction provides rate $E(N-2)/(E+2)N = \dfrac{1-2/N}{1+2/E}$ quantum single deletion error-correcting codes.

\subsection{Classical Erasure Error-Correcting Codes to Classical Deletion Error-Correcting Codes}\label{sec:ecTodc}

\begin{Definition}[Alternating Sandwich Mapping, $F$]\label{def:F}
Let $\Sigma$ be a $2^E$-ary alphabet
and $\beta : \Sigma \rightarrow \{0, 1\}^{E}$ an injection. 
Define $f : \Sigma \rightarrow \{0, 1\}^{E+2t}$ as
$$
f (a) := \bm{1}_{(t)} \bm{a} \bm{0}_{(t)}
$$
where 
$\bm{0}_{(t)}$ is the $t$-repetition of the bit $0$, i.e., $00\dots0 \in \{0,1\}^t$,
$\bm{1}_{(t)}$ is the $t$-repetition of the bit $1$, i.e., $11\dots1 \in \{0,1\}^t$,
and
$\bm{a} := \beta(a)$.
Additionally, define the following map $F : \Sigma^{N} \rightarrow \{0,1\}^{N(E+2t)}$
as
$$
 F(a_1 a_2 \dots a_{N}) := f(a_1) f(a_2) \dots f(a_{N})
$$
as the concatenation of $f (a_i)$.
\end{Definition}

Here,
we observe the image $F( \Sigma^N )$ and $\mathcal{X}$.
By the definition of the alternating sandwich mapping $F$,
any element $\bm{x}$ of $F( \Sigma^N )$ has the following form:
$$
x_1 x_2 \dots x_{(E+2t)N} = \bm{1}_{(t)} *** \bm{0}_{(t)} \bm{1}_{(t)} *** \bm{0}_{(t)}  \cdots \bm{1}_{(t)} *** \bm{0}_{(t)}.
$$
In particular, 
\begin{align}
x_{(E+2t) r + 1}      &= 1, \label{eq:10r1}\\
x_{(E+2t) r + (E+2t)} &= 0, \label{eq:10r10}\\
x_{(E+2t)s} & \neq x_{(E+2t)s+1},  \label{eq:5s}
\end{align}
for $0 \le r < N$ and $1 \le s < N$.

\begin{Lemma}\label{lem:FS_SingleDelCode}
If $S$ is a classical $t$-erasure error-correcting code over a $2^E$-ary alphabet of length $N$,
then $C := F(S) $ is a classical $t$-deletion binary error-correcting code,
equivalently,
$d_L( \bm{x}, \bm{y} ) \ge 2t+2$ for any distinct $\bm{x}, \bm{y} \in C$.

In particular,
the code rate of the binary code $C$ is $\frac{\log_2 |S|}{(E+2t)N}$.
\end{Lemma}

\subsection{Instance of High Rate Quantum Single Deletion Error-Correcting Codes}\label{sec:InstHRC}
Let us focus on the case $t=1$ and the following set $S$.
$$
S 
:= \{ a_1 a_2 \dots a_N \in \mathbb{Z}_{2^E}^N 
     \mid
      \sum_{1 \le i \le N} a_i = 0
   \},
$$
where
$\mathbb{Z}_{2^E}$ is the quotient ring of integers with $2^E$ elements
and $N$ is the multiple of $2^E$.
The set $S$ is an additive single parity-check code
over $\mathbb{Z}_{2^E} = \{ 0, 1, \dots, 2^{E}-1 \}$
with the minimum Hamming distance $2$
and the cardinality $(2^{E})^{N-1}=2^{E(N-1)}$.
Set the injection $\beta : \mathbb{Z}_{2^E} \rightarrow \{ 0, 1\}^E$ as
a map of $E$-digit binary expression.
By Lemma \ref{lem:FS_SingleDelCode},
the image $C := F(S)$ is a classical single deletion error-correcting code,
i.e., a $(t=1)$-deletion error-correcting code.

The image $C$ is divided into $(2^{E})^{N-2}$ sets $\{ X^{(\bm{a})} \mid \bm{a} \in S \}$,
here
$$
X^{(\bm{a})} := \{ F(\bm{a} + \bm{i}_{(N)}) \mid i \in \mathbb{Z}_{2^E} \},
$$
where $\bm{i}_{(N)}$ is the $N$-dimensional vector with all entries $i$, i.e.,
$ \bm{i}_{(N)} = iii \dots i \in (\mathbb{Z}_{2^E})^N$.
Note that $\bm{i}_{(N)} \in S$.
Remark that a set $\{ \bm{a} + \bm{i}_{(N)} \mid i \in \mathbb{Z}_{2^E} \}$
consists of $2^E$ elements and the $j$th entries of the set is $\mathbb{Z}_{2^E}$
for any $1 \le j \le N$.

Each $X^{(\bm{a})}$ consists of just $2^E$ sequences.
Note that $X^{(\bm{a})}$ is a subset of $C$ for any $\bm{a} \in S$,
since $N$ is a multiple of $2^E$ and $\bm{i}_{(N)} \in S$.
Thus $\{ X^{( \bm{a} )} \subset C \mid \bm{a} \in S \}$ is a partition of the classical single-deletion code $C$.
Then a family set $\mathcal{X}$ is defined as a partition of the image $C$:
$$
\mathcal{X} := \{  X^{(\bm{a})} \subset C \mid \bm{a} \in S \}.
$$

\begin{Lemma}\label{lem:calX_RSs}
The partition $\mathcal{X}$ above of the classical single deletion error-correcting code $C$ is homogeneous.
\end{Lemma}

Hence $\mathcal{X}$ is homogeneous by Lemma \ref{lem:calX_RSs}.
We obtain a quantum code from Corollary \ref{cor:corOfMain}
and Theorem \ref{thm:mainGeneralConstruction}.

\begin{Theorem}\label{thm:highRateExistence}
Let $E$ be a positive integer, $N$ a positive multiple of $2^E$, and $\mathcal{X}$ a family set
from the classical single deletion error-correcting code $C$ constructed as above.
Then a quantum single deletion error-correcting code is obtained from $\mathcal{X}$
Its code length is $(E+2)N$ and its complex dimension is $| \mathcal{X} | = (2^E)^{N-2} = 2^{E(N-2)}$.
Therefore, the code rate is $E(N-2)/(E+2)N = \dfrac{1-2/N}{1+2/E}$.

Hence, for any $0 < R < 1$,
there exists a quantum single deletion error-correcting code
whose rate is greater than $R$.
\end{Theorem}

\section{Concluding Remarks}\label{sec:concRem}
The research results described in this paper can be roughly summarized as follows.
\begin{itemize}
\item The three conditions \cratio, \coutd, and \cind were extended so that the dimension of the quantum error-correcting codes can be defined flexibly.
\item In order to deal with the three conditions, the properties of the objects described as $\Delta_{I,b}(X)$ and $X_{I,b}$ were investigated.
\item The probabilistic aspect of $\lambda_{I,b}$ appearing in \cratio was obtained.
\item The relationship between classical deletion error correction and the three conditions were investigated.
\item The $b$-run support and BRS stable were introduced. The author thinks that these are easier to handle than the three conditions.
\item Homogeneous partition is introduced. The homogeneous partition of the classical error-correcting code is shown to satisfy the three conditions. This allowed us to propose a method to construct a quantum deletion code from a classical deletion error correcting code.
\item The homogeneous partition was concretely constructed. In particular, it was shown that a quantum single deletion error-correcting code with a high code rate could be constructed.
\end{itemize}

This paper aimed to contribute deletion error-correcting code construction and fundamental theory on quantum deletion error-correcting codes.
We have extended the previous three conditions to increase the dimension of quantum codes.
In addition, the author proposed a new approach called homogeneous partition.
This approach has become a bridge between classical and quantum deletion error-correcting codes.
The author hopes that the homogeneous partition will be used as a catalyst to stimulate research on both deletion codes.
The homogeneous partition is sufficient but not necessary to the three conditions.
Nevertheless, we were able to construct a quantum code with a high coding rate.
The author believes this is because it is easier to use the classical deletion codes than the three conditions.

There are many unsolved problems in the study of quantum deletion codes.
The following are examples of problems.

\begin{enumerate}
\item[Q1:] Can we define the quantum Levenshtein distance?

Levenshtein distance is very useful in the study of classical deletion codes.
The distance is widely used in not only code theory but also linguistics \cite{jan2007receptive}, computer science \cite{hirschberg1975linear,wagner1974string} and etc.
With the development of classical information science, Levenshtein distance is used for research for racetrack memories and DNA storages recently.
With the development of quantum information science, the quantum Levenshtein distance may become significant.

\item[Q2:] How does ordering affect the combination of quantum insertion and quantum deletion errors?

It was mentioned in this paper that any single unitary error could be realized by a combination of a single-quantum deletion error followed by a single-quantum insertion error.
However, this is no longer true if the order of the combinations is reversed.
This can be explained as follows.
\cite{shibayama2022equivalence} shows that the insertion error is limited when the quantum state is pure.
In other words, if we say $| \phi \rangle \langle \phi |$ as the state before insertion, the state after insertion can only be the state with permutation on $| \phi \rangle \langle \phi | \otimes \sigma $
for some quantum state $\sigma$, i.e. the pure state $| \phi \rangle \langle \phi |$ and the inserted state $\sigma$ are separable.
The state is no longer pure if a deletion error at a position equivalent to $| \phi \rangle \langle \phi |$ occurs.
Since a unitary error transfers a pure state to a pure state, a combination of insertion followed by deletion cannot be a unitary error.

In classical information theory, the ordering does not matter in the combination of insertion and deletion errors.
Any error by first a deletion error and then an insertion error
can be made by  first an insertion error and then a deletion error.
The converse is also true.

The author is interested in the combination of insertion and deletion errors that can make a difference between classical and quantum information.

\item[Q3:] Can the homogeneous partition be generalized for multi-deletion error-correction?
Moreover, can quantum multi-deletion error-correcting codes with a high code rate be realized from the generalization?

\item[Q4:] Can quantum deletion error-correcting codes correct quantum insertion errors?
Conversely, can quantum insertion error-correcting codes correct quantum deletion errors?
In other words, are quantum deletion error-correctable and quantum insertion error-correctable equivalent?

In classical coding theory, equivalence between deletion error-correctable and insertion error-correctable is known \cite{levenshtein1966binary}.
Does a similar equivalence hold in quantum code theory?
This question is related to Q1 and Q2.
However, I would like to suggest that equivalence may not be derived immediately from Q1 and Q2.
This is because, with quantum information, operations are limited to those that are quantum mechanically feasible.
In other words, when discussing equivalence, one must also include that the operation of deletion error-correction is quantum mechanically feasible, and the operation of insertion error-correction is quantum mechanically feasible.
This argument was not necessary for classical codes.
\end{enumerate}


\section*{Acknowledgments}
This paper is partially supported by
KAKENHI 21H03393.

\bibliographystyle{plain}
\bibliography{reference}

\end{document}